\documentclass[pra,amsmath,amssymb,superscriptaddress,showpacs,notitlepage,nofootinbib]{revtex4}
\usepackage{graphics}
\usepackage{graphicx}
\usepackage{epsfig}
\usepackage{bbm}
\usepackage{amssymb}
\usepackage[latin1]{inputen c} 
\usepackage[usenames]{color}

\newcommand{\be}{\begin{equation}}
\newcommand{\ee}{\end{equation}}
\newcommand{\eea}{\end{eqnarray}}

\newcommand{\va}[1]{\ensuremath{(\Delta#1)^2}}

\newcommand{\ex}[1]{\ensuremath{\left\langle{#1}\right\rangle}}
\newcommand{\exs}[1]{\ensuremath{\langle{#1}\rangle}}

\newcommand{\ket}[1]{\ensuremath{|#1\rangle}}

\newcommand{\bra}[1]{\ensuremath{\langle#1|}}

\newcommand{\braket}[2]{\ensuremath{\langle #1|#2\rangle}}

\newcommand{\kommentar}[1]{}

\newcommand{\EQ}[1]{(\ref{#1})}
\newcommand{\EQEQ}[1]{Equation~(\ref{#1})}
\newcommand{\REF}{}
\newcommand{\REFS}{}

\newcommand{\komment}[1]{}
\renewcommand{\prl}{{\it Phys. Rev. Lett. }}
\renewcommand{\pra}{{\it Phys. Rev. A }}
\renewcommand{\prb}{{\it Phys. Rev. B }}
\newcommand{\nature}{{\it Nature }}
\newcommand{\naturephys}{{\it Nature Phys. }}
\newcommand{\science}{{\it Science }}
\renewcommand{\preprint}{{\it Preprint }}

\newcommand{\fl}{}
\newcommand{\ack}{\section*{Acknowledgments}}

\newif\ifdraft

\drafttrue

\ifdraft
  \newcommand{\otext}[1]{{\color{green}#1}}
  \newcommand{\atext}[1]{{\color{Orange}#1}}
\else
  \newcommand{\otext}[1]{}
  \newcommand{\atext}[1]{}
\fi

\begin{document}
\title{Generation of macroscopic singlet states in atomic ensembles}
\date{\today}
\begin{abstract}
We study squeezing of the spin uncertainties by quantum non-demolition (QND) measurement
in non-polarized spin ensembles. Unlike the case of polarized
ensembles, the QND measurements can be performed with negligible
back-action, which allows, in principle, perfect spin
squeezing as quantified by [G. T\'{o}th {\em et al.}, Phys. Rev.
Lett. 99, 250405 (2007)]. The generated spin states approach
many-body singlet states, and contain a macroscopic number of
entangled particles, even when individual spin is large. We introduce the Gaussian treatment of unpolarized spin states and use it to
estimate the achievable spin squeezing for realistic experimental
parameters. Our proposal might have applications for magnetometry with a high spatial resolution
or quantum
memories storing information in decoherence free subspaces.
\end{abstract}

\author{G\'eza T\'oth}
\email{toth@alumni.nd.edu}
\affiliation{Department of Theoretical Physics, The University of the Basque Country,\\
P.O. Box 644, E-48080 Bilbao, Spain }
\affiliation{IKERBASQUE, Basque Foundation for Science, E-48011
Bilbao, Spain}
\affiliation{Research Institute
for Solid State Physics and Optics, \\Hungarian Academy of Sciences,\\
P.O. Box 49, H-1525 Budapest, Hungary}

\author{Morgan W. Mitchell}
\affiliation{ICFO-Institut de Ciencies Fotoniques, Mediterranean
Technology Park, \\ E-08860 Castelldefels (Barcelona), Spain}


\maketitle



\section{Introduction}

Realization of large coherent quantum systems exhibiting
nonclassical behavior is at the center of attention in many-body
quantum experiments with cold atoms \cite{HS99} and ions
\cite{VR01}. In a system of spin-$\frac{1}{2}$ particles spin
squeezing \cite{K93,K93b} is one of the most successful approaches for
creating large scale quantum entanglement
\footnote{It is instructive to see the relation of spin squeezing to polarization squeezing \cite{LK06,LK06b}.}. In a spin squeezed state,
the collective spin of an ensemble of particles is almost completely
set into one direction, while the variance of an orthogonal spin
component is decreased below the standard quantum limit. In the
special case of spin-1/2 atoms this implies atom-atom entanglement
\cite{SD01} \footnote{For $j>1/2,$ spin squeezing can be also the result
of entanglement between the spin-$1/2$ constituents of the particles, e.g.,
entanglement between the nuclear and electronic spins \cite{F08}.}. In a Quantum Non-Demolition (QND) scheme, the particles
interact with a light field, which is subsequently measured
projecting the atoms into a squeezed state \cite{KB98,KB98b}. Typically,
the length of the squeezing dynamics is limited by losses, and a
short-time approximation can be applied.  In this
regime, two of the spin components orthogonal to the large mean spin
behave like the canonical variables $x$ and $p.$ The Gaussian
formalism can then be used for modeling \cite{GC02}, and can include
realistic effects of noise and imperfections
\cite{book,MM04,HM04,EM05,KM08}.

At this point the questions arise: Is it possible to realize a
protocol for the creation of entanglement by squeezing the spin
uncertainties that would also work for higher spins? This is
important since most experiments use atoms with larger spins. The
solution is not easy: Known methods for creating spin-$\frac{1}{2}$
entanglement by spin-squeezing cannot straightforwardly be
generalized to higher spins, without restricting dynamics or the
detection to a spin-$1/2$ subspace \cite{DC02,MZ02}. Moreover, from
the point of view of modeling spin systems, one might ask: Is it
possible to extend the Gaussian formalism to unpolarized spin
states? This is important if we want to depart from the usual setups
with large collective spin.

In this paper, we show how to create and detect entanglement by QND
measurement of collective spin in an unpolarized ensemble.
We show how to create a many-body singlet state of atoms without requiring that
they interact with each other and the system settles in a ground state of some
antiferromagnetic Hamiltonian.
Our proposal works even for particles with a large spin. With realistic
experimental parameters for $^{87}$Rb \cite{EM05}, the predicted
squeezing dynamics are robust to decoherence, and produce a
many-atom singlet state. Unlike standard spin squeezing, the method
creates entanglement even in the limit of very strong interaction,
which might be used in experimental implementations with cavities
\cite{NM08,SL08}, or in any multi-atomic system in which a von Neumann
measurement of the collective spin is possible. We demonstrate the
validity of the Gaussian approximation for unpolarized spin states.
For the lossless case, we confirm our finding with comparison to the
exact model.

The paper is organized as follows. In Section 2, we present the spin squeezing parameter to detect
the entanglement of many-body singlet states, and also discuss the properties of the singlets we aim to prepare.
In Section 3, we describe the squeezing process. First, we consider the lossless case and present a model
based on a Gaussian approximation. Later, we include
decoherence in the model. For the lossless case,
we compare our results to the results of the exact model. In the Appendix,
we present the deltails of the calculations for the exact model.

\section{Detecting the entanglement of singlet states}

In this paper, we will use the generalized spin squeezing parameter
\begin{equation}
\xi_s^2:=\frac{\va{J_x}+\va{J_y}+\va{J_z}}{J}, \label{Jxyz}
\end{equation}
where $J_l$ are the components of the collective angular momentum, $\va{J_l}=\exs{J_l^2}-\exs{J_l}^2$
and for a system of $N$ spin-$j$ particles we define $J:=Nj.$ It has already been
shown in \REFS\cite{T04,W05,TK07,TK07b,GT09} that any
state giving $\xi_s<1$ is entangled (i.e., not fully separable).
For completeness, we present briefly the proof for
\EQ{Jxyz}. For pure product states of the form $\ket{\Psi_{\rm p}}=\otimes_{k=1}^N \ket{\psi_k},$
we have
\be
\sum_{l=x,y,z} \va{J_l}=\sum_{l=x,y,z} \sum_{k=1}^N \va{j^{(k)}_l}_{\ket{\psi_k}}\ge Nj,
\label{crit}
\ee
where $ j^{(k)}_l$ denotes the spin coordinates of particle $(k)$ for $l=x,y,z.$
Here we used the fact that $\sum_l \va{j^{(k)}_l}_{\ket{\psi_k}}\ge j.$
For a mixture of pure product states, i.e., for separable states, \EQ{crit} remains true
since the variance is concave in the state.

\begin{figure}
\centerline{ \epsfxsize15cm \epsffile{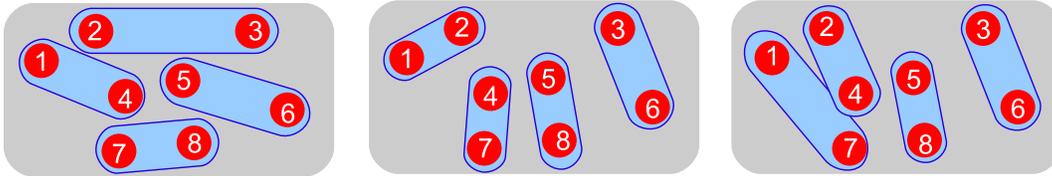}}
\caption{
For qubits, the permutationally invariant singlet is an equal mixture of all possible arrangements of two-particle singlets.
Three of such arrangements are shown for eight particles. Note that the eight atoms are arranged in the same way on the figures,
only the pairings are different.
 }
 \label{singlet}
\end{figure}

The states giving $\xi_s=0$ are called many-body
singlet states \cite{C03}. In particular, an equal mixture of all
pure singlets, expected to arise in permutationally invariant
systems, has intriguing entanglement properties \cite{T04,TK07,TK07b}.
The bipartite entanglement of this state has already been determined for qubits \cite{LT05}. It
is very mixed, yet its entanglement is robust to noise \cite{GT09}.
For qubits, the singlet state studied
in this paper is an equal mixture of all states composed of
two-qubit singlets as can be seen in figure~\ref{singlet}.

For an imperfect realization, $N\xi_s^2$ gives an upper
bound on the number of particles unentangled with other particles
\cite{TK07,TK07b}. This can be seen as follows. Let us consider a pure state
of the form $\otimes_{k=1}^M \ket{\psi_k} \otimes \ket{\psi_{M+1,...,N}},$
which have $M$ particles unentangled with the rest. For such a state,
we have
\be
\va{J_l}=\sum_{k=1}^M \va{j^{(k)}_l}_{\ket{\psi_k}}+
[\Delta (\sum_{k=M+1}^N j^{(k)}_l)]^2_{ \ket{\psi_{M+1,...,N}}}.
\ee
Hence,
\be \fl
\va{J_x}+\va{J_y}+\va{J_z}\ge \sum_{k=1}^M \va{j^{(k)}_x}_{\ket{\psi_k}}+\va{j^{(k)}_y}_{\ket{\psi_k}}+\va{j^{(k)}_z}_{\ket{\psi_k}}\ge Mj.
\ee
Due to the concavity of the variance, mixing pure states with $M$ or more unentangled particles,
we still have $N\xi_s^2\ge M.$
If $N\xi_s^2$ for some quantum state is smaller than this bound, then the state cannot be obtained
by preparing pure states having at least $M$ unentangled spins and mixing them.
If $N(1-\xi_s^2)$ is a large number then we can say that entanglement between macroscopic
number of particles is present in the system in this sense.

Next, we will review the relevant properties of singlet states.
Many-body singlet states are at the center of attention in condensed
matter physics, and also in other areas of quantum physics. They
appear as ground states of many fundamental spin models.
There have
been numerous experiments and experimental proposals for studying
such states \cite{EI08,EI08bb,EI08b,EI08c}.
Surprisingly, the permutationally invariant singlet
appears even in quantum gravity calculations of black hole entropy
\cite{LT05}. Its realization is difficult as a $T=0$ thermal state of a spin system, since in this case the
system Hamiltonian is $J_x^2+J_y^2+J_z^2,$ which is essentially the sum of two-body interactions
connecting {\it all} spin pairs.

Singlets play an important role in quantum information processing.
They are invariant under
transformations of the form
$U_\beta=\exp(-i\beta J_{\vec{n}}),$ which describe the effect of an external homogenous magnetic field
on the spins and $\vec{n}$ is the direction of the field. $\xi_s$ can directly be related
to the decrease of fidelity of a pure state $\ket{\Psi}$ under an external magnetic field, defined as
\be \Delta F:= 1-\vert \bra{\Psi}U_\beta \ket{\Psi} \vert^2.\ee Straightforward calculations show that
\begin{equation}
\Delta F \approx \va{J_{\vec{n}}} \beta^2 \le \beta^2 J\xi_s^2
\end{equation}
for small $\beta$  \footnote{For mixed states, $4J\xi_s^2=4\sum_l \va{J_l}$ bounds from above the Quantum Fisher Information $F_Q[\rho,J_{\vec{n}}],$
which gives the maximum phase
estimation sensitivity in an interferometer using the quantum state $\rho$
and unitary dynamics $\exp(-iJ_{\vec{n}}t).$ Here we used the fact that $F_Q[\rho,J_{\vec{n}}]\le 4\va{J_{\vec{n}}}$
\cite{PS09}.}.
Thus, small $\xi_s$ indicates insensitivity to external fields, which
makes singlet states
applicable to encoding quantum information in
decoherence free subspaces \cite{LC98}, for sending
information independent of a reference direction \cite{BR03},
or possibly, for metrological applications
in which insensitivity to external
homogenous magnetic fields is needed, e.g., measuring the spatial gradient or
fluctuations of the fields. Note that measuring field gradients
is possible with two almost completely polarized atomic ensembles in an entangled state \cite{WJ09,EH08}.
A bipartite singlet can also be used for this aim
\cite{CD09}. However, using
the singlet of a {\it single} atomic ensemble can lead to magnetometry
with a better spatial resolution.

\section{Decreasing the collective spin uncertainties}

\subsection{Outline of the squeezing procedure}

In this paper,
we consider a procedure to produce an atomic state with $\xi_s<1$
from a non-entangled state with
$\xi_s\sim 1.$ For the initial state and throughout the dynamics we have
$\left<\bf J\right> = 0.$
Suitable initial states include the thermal state (completely mixed state), for which
$\xi_s=\sqrt{j+1}$, and pure product states of the form
$\ket{\Psi_{\rm p}}=\otimes_k \ket{\psi_k},$ with
$\vert\exs{\vec{j}}_{\psi_k}\vert=j$ for which
$\xi_s=1.$  Spin coherent states also have $\xi_s=1,$  however, for them $\left<\bf J\right> \ne 0.$
After the initial state was created,
$\xi_s$ is reduced by sequentially measuring and if necessary correcting by
feedback the spin components.

Next, we will examine the dynamics of the expectation values and variances of
collective spin operators and we will determine how they depend on the particle
number $N.$ In particular, $A\sim N^p$ will denote that the quantity $A$ is proportional to $N^p$ for large $N.$

The first squeezing step is a QND measurement of the atomic spin component $J_x.$
The realization is similar to QND measurements in polarized atomic ensembles \cite{KB98,KB98b}.
The atoms interact with light, which is subsequently measured.
As discussed later, this procedure can reduce the value of $\va{J_x}$ considerably
 but with only a minor increase in $\va{J_y}$
and $\va{J_z}$. However, the procedure results in nonzero $\ex{J_x}.$
Fortunately, the expectation value of the spin component $J_x$ remains small and we obtain
$\vert\ex{J_x}\vert\lesssim\sqrt{N}$.
Feedback, using the data coming from the measurement of the light can then be applied to restore
$\exs{J_x}=0.$
 This feedback step is similar to the coherent rotation
in polarized spin-squeezing \cite{feedback1,feedback2}, but does not require a
large average polarization.
However, it is now not simply a rotation, but an incoherent process.
The feedback introduces negligible noise:
It can be shown that for making
$\ex{J_x}$ zero, $\sim \sqrt{N}$ atoms are affected,
introducing extra noise $\va{J_x}\sim \sqrt{N},$  which negligible relative to the initial noise in the large $N$ limit. \footnote{
Scattering sets the two-body correlations $\exs{j_l^{(m)}j_l^{(n)}}$ to zero, where $m$ is one of the spins
that went through scattering, while $n$ in one of the non-affected spins. From $\exs{J_l^2}\ge 0,$ a lower bound on
the average two-point correlations is $\sim - \tfrac{1}{N}.$ Since $\sim\sqrt{N}$ atoms are affected by scattering,
the number of two-point correlations set to zero by scattering is $\sim N\sqrt{N}.$ Thus,
the increase
of $\va{J_l}$ is at most $\sim\sqrt{N}.$}. Alternately, instead of feedback,
post-selection could be used to identify cases with low $\ex{J_x}$.

If the initial state was a pure product state of the type mentioned above, measurement of
$J_x$ is sufficient to produce a squeezed state with $\xi_s<1.$
For the thermal state, further squeezing steps are needed
for  $J_y$ and
$J_z$. Since $\ex{\bf J}=0$, the generalized uncertainty
relations $\va{J_k} \va{J_l} \ge \tfrac{\hbar^2}{4} \ex{J_m}^2$ do not
enforce a measurement back-action.  As such, the QND measurements
can produce a state with $\va{J_x},\va{J_y},\va{J_z}$ all significantly reduced
 so that $\xi_s^2 \ll 1.$

\subsection{Lossless case}

We now describe the details of our calculations.
We employ
methods  developed
for treating the Gaussian dynamics of continuous variable systems \cite{KB98,KB98b,GC02,HM04},
which we adapt to the case of unpolarized ensembles of spin-$j$ particles. First, the QND
measurement is modeled in the absence of decoherence, which allows comparison with an
exact calculation. Then, we incorporate decoherence due to light scattering
\cite{MM04}.

For modeling the QND pulse, the atoms are described by the
$J_l$ operators, while the light
pulse is characterized by the Stokes operators $S_l$ \cite{MM04,KM08}.
We choose
the initial state to be the completely mixed atomic state,
 $\varrho_0:=\tfrac{1}{(2j+1)^N}\openone$ 
and a fully-polarized optical state with $\ex{\bf S}=(S_0,0,0).$
The full system is described by the operators
\be R=\left\{\tfrac{J_x}{\sqrt{J}},\tfrac{J_y}{\sqrt{J}},\tfrac{J_z}{\sqrt{J}},
\tfrac{S_x}{\sqrt{S_0}},\tfrac{S_y}{\sqrt{S_0}},\tfrac{S_z}{\sqrt{S_0}}\right\}\ee
with a covariance matrix
\be\Gamma_{mn}:=\tfrac{1}{2}\exs{R_mR_n+R_nR_m}-\exs{R_m}\exs{R_n}.\ee
As shown by simple calculations,
for large $N$ the initial state is Gaussian for the $R_k$ operators.
That is, symmetric moments with order higher than second
can be obtained from lower order ones according to the theory
of Gaussian distributions, knowing that cumulants with order three and higher are zero \cite{distbook}.
In other words,  concerning the moments of $R_k,$
the state is completely characterized by $\Gamma$, $\exs{\bf S}$ and $\exs{\bf J}.$

The first step of the QND measurement of $J_x$
is interaction between the atoms and the light
 via the Hamiltonian
\be H=\hbar\Omega J_x S_z.\ee This suggests a characteristic time-scale \cite{book}
 \be\tau:=\tfrac{1}{\Omega\sqrt{S_0J}}.\label{tau}\ee
 The dynamical equations of $\Gamma_{mn}$ can be obtained from the Heisenberg equation of
motion for the operators $R_k$ given as
\be R_k^{({\rm out})}=R_k^{({\rm in})} -it[R_k^{({\rm in})},H],\ee with $\hbar =1.$
For example,  the dynamics of $R_5$ is obtained as
\be
R_{5}^{\rm (out)}=R_{5}^{\rm (in)}+\tfrac{\kappa}{\sqrt{S_{0}}}R_{4}^{\rm (in)}R_{1}^{\rm (in)}.
\ee
where the coupling constant is defined as $\kappa:=\frac{t}{\tau}.$ Hence, for the dynamics of the variance of $R_5$ we obtain
\begin{eqnarray}
\fl \exs{(R_{5}^{\rm (out)})^{2}}=\exs{(R_{5}^{\rm (in)})^{2}}
+\tfrac{\kappa^2}{S_{0}}\exs{(R_{4}^{\rm (in)})^2(R_{1}^{\rm (in)})^2}+\tfrac{\kappa}{\sqrt{S_{0}}}\exs{R_{1}^{\rm (in)}\{R_{4}^{\rm (in)},R_{5}^{\rm (in)}\}_+},
\end{eqnarray}
where $\{A,B\}_+$ is the anticommutator of $A$ and $B.$
Knowing that due to symmetries of the setup for all times $\ex{R_k}=0$ for $k=1,2,3,5,6,$ we obtain
\begin{eqnarray}
\fl\exs{(\Delta R_{5}^{\rm (out)})^{2}}&=&\exs{(\Delta R_{5}^{\rm (in)})^{2}}
+\tfrac{\kappa^2}{S_{0}}\exs{(\Delta R_{4}^{\rm (in)})^{2}(\Delta R_{1}^{\rm (in)})^{2}}
+\tfrac{\kappa^2}{S_{0}}\exs{R_{4}^{\rm (in)}}^{2}\exs{(\Delta R_{1}^{\rm (in)})^{2}}\nonumber\\
&+&\tfrac{\kappa}{\sqrt{S_{0}}}\exs{\Delta R_{1}^{\rm (in)}\{\Delta R_{4}^{\rm (in)},\Delta R_{5}^{\rm (in)}\}_+}
+\tfrac{2\kappa}{\sqrt{S_{0}}}\exs{\Delta R_{1}^{\rm (in)}\Delta R_{5}^{\rm (in)}}\exs{R_{4}^{\rm (in)}},\nonumber\\\label{R5}
\end{eqnarray}
where we used the notation $\Delta R_k:=R_k-\ex{R_k}.$

Let us now consider dynamics for $t\lesssim\tau.$ Knowing that $\ex{R_4}=\sqrt{S_0}$ for $t=0$ and $\ex{\Delta R_k \Delta R_l \Delta R_m \cdot \cdot \cdot }\lesssim 1,$ we can examine how the different terms depend on $S_0.$
We find that on the right hand side of \EQ{R5} the second term is of order $\tfrac{1}{S_0},$
the fourth term is of the order $\tfrac{1}{\sqrt{S_0}},$ while the rest of the term are of order $1.$
Thus, assuming a large number of photons, that is, a large $S_0,$
several terms can be neglected and we obtain
\begin{eqnarray}\fl
\exs{(\Delta R_{5}^{\rm (out)})^{2}}=\exs{(\Delta R_{5}^{\rm (in)})^{2}}
+\tfrac{\kappa^2}{S_{0}}\exs{R_{4}^{\rm (in)}}^{2}\exs{(\Delta R_{1}^{\rm (in)})^{2}}
+\tfrac{2\kappa}{\sqrt{S_{0}}}\exs{\Delta R_{1}^{\rm (in)}\Delta R_{5}^{\rm (in)}}\exs{R_{4}^{\rm (in)}}.
\end{eqnarray}
Long, but straightforward calculations show that several terms can also be neglected in the dynamical equation for variances of the other $R_k$ variables, assuming large $J$ and $S_0.$
Hence, for the evolution of the covariance matrix
\be\Gamma_P=M \Gamma_0 M^T\ee is obtained, where
$M$ is identity matrix, apart from
$M_{5,1}=\frac{\exs{R_4^ {\rm (in)}}}{\sqrt{S_0}}\kappa.$
The evolution of the expectation values
is described as \be\exs{R_k^{({\rm out})}}=\sum_{l} M_{kl}\exs{R_l^{({\rm in})}}.\ee
Similar analysis shows that the dynamical equations for higher order moments
of $R_k$ can also be simplified if $t\lesssim\tau.$ When computing the dynamics
of these moments, instead of the Heisenberg equation of motion,
$R_k^{({\rm out})}=\sum_l M_{kl} R_l^{({\rm in})}$ can be used. Under this dynamics, which is a linear
mapping between operators, the state remains Gaussian.
Note, however, that this
approximation breaks down for much larger times $t\sim\tau\sqrt{J}.$ (See also Appendix.)

The second step of the QND process is
measuring $S_y$ of the light. The theory of Gaussian states can be used to describe the measurement,
which
is modeled with a projection \cite{tobepub} \footnote{\EQEQ{Gamma_M}
is analogous to the formula for the behavior of the correlation matrix during a von Neumann measurement in
the case of Gaussian continuous variable systems described in \cite{GC02}. However,
\EQ{Gamma_M} is based on the theory of $SU(2)$ Wigner functions \cite{spinwigner} rather than on the theory
of Wigner functions for multimode systems. For polarized ensembles, \EQ{Gamma_M} has already been used in \REF\cite{KM08}.}
 \be
\Gamma_{M} =\Gamma_P-\Gamma_P(P_y\Gamma_P P_y)^{\rm MP}\Gamma_P^T.\label{Gamma_M}
 \ee
Here ${\rm MP}$ denotes the Moore-Penrose pseudoinverse and $P_y ={\rm diag}(0,0,0,0,1,0)$.  As
described above, the QND measurement minimally disturbs the
unmeasured components. In the Gaussian approximation the variance
of the other spin components remain unchanged.

Measuring the other
$J_k$ components is analogous to the $J_x$ case. Note that magnetic fields
could be used to rotate the collective spin to facilitate the measurement of the different
spin components. Finally, as we have already noted, after each squeezing step,
feedback has to be used to restore the zero expectation value of
the angular momentum coordinates. We have also mentioned that
post-selection can be used in the place of
feedback. In this case a feedback scheme does not have to be realized,
however, part of the experiments must be discarded.
Thus, when a QND measurement determines $\exs{J_l}$, only
cases with $|\exs{J_l}| \le B$ are retained. We define
$I(f(x),L):=\int_{-L}^L f(x) \exp(-x^2/2\Delta^2)dx,$ where $\Delta$
is the width of the distribution of values obtained when measuring $\exs{J_l}.$  The effect of
post-selection is given as ${\rm var}(\exs{J_l})_{\rm after}=\mu {\rm var}(\exs{J_l})_{\rm
before}$ where $\mu:=[I(x^2,B)/I(1,B)]/[I(x^2,\infty)/I(1,\infty)]$ and the fraction of
post-selected experiments is $q:=I(1,B)/I(1,\infty).$  Hence, for a
moderate post-selection of $B/\Delta=0.678$ we get $q=0.5$ for all the three
squeezing steps, and the variances decrease to $14\%$ of their
original values. For the case of $B/\Delta=1.150,$ we get $q=0.75$ and the decrease is  $37\%.$

Let us now make the calculations for realistic parameters.
We
consider $N=10^6$ $^{87}$Rb atoms with spin $j=1,$ and for the light
field $S_0=5\times10^7.$
Sequential squeezing of the $x,$ $y$ and $z$ spin
components is shown in figure~\ref{J2xyz}. The horizontal axis indicates the
total interaction time, with successive intervals of up to $2 \tau$ for measurement
of $J_x$, $J_y$, and $J_z$, respectively.  Results are shown for squeezing from a
thermal state and also from
 \be
 \ket{\Psi}_0' :=
 \ket{+j}^{\otimes \tfrac{N}{2}}\otimes \ket{-j}^{\otimes \tfrac{N}{2}}. \label{init2}
 \ee
We obtain $\xi_s^2=0.32$ and $\xi_s^2=0.20$ for the completely mixed initial atomic state and for
\EQ{init2}, respectively.
Remarkably, the QND interaction can be solved exactly for the initial state \EQ{init2}, as shown in
the Appendix. The results are presented in figure~\ref{J2xyz}.
In that calculation  we find that, for large $N$, time
$t\sim\tau\times J^{0.25}$ gives $\va{J_x}\sim
\sqrt{J}$ and the two halves of the atoms
remain almost fully polarized into the $+z$ and $-z$ directions, respectively.
For much longer times, the two halves are not fully polarized any more.
In particular,
for $t\sim\tau_2:=\tau\times \sqrt{J}$ we obtain $\va{J_x}\sim
1,$ and $\xi_s^2=\tfrac{1}{2}.$
Thus, we have squeezing even in the long-time
(von Neumann) limit.

\subsection{Model including losses}

We now incorporate decoherence, following ideas from \REFS\cite{MM04,HM04,KM08},
adapted to our use of a correlation matrix
of all the three spin components. In particular, a
parameter $\eta$ describes the probability that an atom suffers spontaneous excitation
due to the off-resonant probe, and thus describes the fraction of atoms that
decohere during the QND process. For simplicity, we assume that the atoms
end up in the completely mixed state, which is usual for handling the effects of noise.
We obtain
 \be
 \Gamma_P'=(\openone-\eta D)M\Gamma_0 M^T(\openone-\eta D)
 +\eta(2-\eta)D\Gamma_{\rm noise},\label{Gamma_Pprime}
 \ee
where $D={\rm diag}(1,1,1,0,0,0)$ and $\Gamma_{\rm noise}={\rm
diag}(1,1,1,0,0,0)\times \frac{n_j}{j}$ \footnote{Compare \EQ{Gamma_Pprime} with (13)  in  \REF{\cite{MM04}}.}. Here $n_j$ is the variance of $j_x$ for a spin-$j$
particle in a totally mixed state.
The decoherence is connected to $\kappa$ through
\be\eta=Q\frac{\kappa^2}{\alpha},\ee where $\alpha$ is the resonant
optical depth of the sample and $Q=1$ for the spin-$\tfrac{1}{2}$ case. For the spin-$1$ case,
$Q=\tfrac{8}{9}$ if the near-resonant intermediate state has $j=0$
\cite{transitions}. Using these techniques, we calculate the
degree of squeezing as a function of the time for different,
experimentally feasible, values of $\alpha$ \cite{HM04,EM05}.

\begin{figure}
\centerline{ \epsfxsize10cm \epsffile{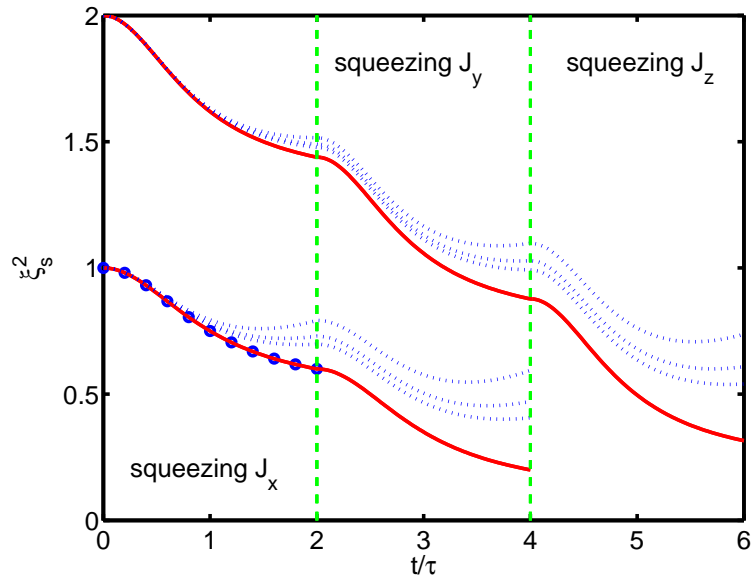}}
\caption{
Dynamics of the spin squeezing parameter $\xi_s^2$ as a
function of the time for $^{87}$Rb atoms with
spin $1.$  Sequential QND measurements are made of the $x, y$ and $z$ spin components, with interaction time $t$ up to $2\tau$.  The initial state is the completely mixed state (upper traces) or the state of \EQ{init2} (lower traces). (Solid) Gaussian approximation without losses. (Dotted) Gaussian approximation including losses with (from top to bottom) $\alpha=50,75,100.$
(Dots) Exact model.
 }
 \label{J2xyz}
\end{figure}

The results, shown in figure~\ref{J2xyz}, indicate that considerable squeezing is indeed possible under
realistic conditions. For the completely mixed atomic initial state,
we obtained $\xi_s^2=0.74,0.61,$ and $0.54$ for $\alpha=50,75$ and $100,$ respectively.
For comparison, polarized spin states have been squeezed
in variance by $\sim50\%$ under similar conditions.  Thus, our proposal is realistic with present-day
technology. The values of $\alpha$ chosen reflect the
state-of-the-art for single-pass optical probing  \cite{KK09,AW09}.  The use of an optical cavity
could boost the effective $\alpha$ by orders of magnitude \cite{SL08}.

\section{Conclusions}

In summary, spin squeezing of unpolarized atomic ensembles by QND
measurement shows several intriguing differences relative to
polarized samples.  The absence of a significant measurement
back-action allows simultaneous squeezing of all spin components
and approaches a macroscopic singlet state.  Spin squeezing of this
type implies entanglement of a macroscopic number of particles, for
arbitrary spin.  To treat this problem, we have extended the Gaussian formalism to
include the dynamics of all the three spin components and their
correlations for an unpolarized ensemble \footnote{Modeling lossy dynamics
is also possible by writing down the infinite hierarchy of  dynamical
equations for many-body correlations and truncate it at second order \cite{CC07}.}.
The advantage of our approach is that it is possible to determine the
area of validity for our model,
and we can also incorporate von Neumann measurements. In the lossless case, the results
agree with an exact calculation.  Realistic calculations including
decoherence indicate that production of
these macroscopic singlet states should be possible with existing
ensemble systems. In the future, it will be interesting to look at
the possibility of storing quantum information in the decoherence free subspace \cite{KM01},
formed by different singlet states, obtained for $j>\tfrac{1}{2}$ when using different initial
states for our squeezing procedure. The method of modeling
large spin systems can be generalized to multiple ensembles
and a series of light pulses or even for modeling
many-particle quantum systems in other
areas of physics.

\ack

We thank A.~Ac\'{\i}n, N.~Brunner, A.~Cere, J.I.~Cirac, G.~Giedke, O.~G\"uhne,
C.~Hadley, K.~Hammerer, M.~Koschorreck, M.~Lewenstein, and M.~Napolitano for
fruitful discussions. We thank the support of the EU (OLAQUI, SCALA,
QICS), the National Research Fund of Hungary OTKA (Contract No.
T049234), the Spanish MEC (Ramon y Cajal Programme, Consolider-Ingenio 2010
project ''QOIT'', project No. FIS2009-12773-C02-02, ILUMA project with Ref. FIS2008-01051).

\appendix

\section{Exact model for squeezing $\va{J_z}$ starting from the state \EQ{init2}}

\subsection{The $j=\tfrac{1}{2}$ case.}
We briefly describe a
method that makes it possible to model exactly the QND process
without using the Gaussian approximation. First, let us consider
the initial state \EQ{init2}
for the $j=\tfrac{1}{2}$ case  and
measure $J_x$ with the QND interaction. We divide the atoms into two groups.
In the first group, initially $N_1$ atoms are in the $\ket{+\tfrac{1}{2}}_z$ state, while
in the second group $N_2$ atoms are in the $\ket{-\tfrac{1}{2}}_z$ state.
We define the angular momentum
operators $J_{k,l}$ with $k=1,2$ corresponding to the two groups.
We choose the two halves equal: $N_1=N_2=\tfrac{N}{2}.$
Moreover, without the loss of generality we choose
$N_k$ to be even since in this case $J_{k,l}$ have integer eigenvalues between $-\tfrac{N_k}{2}$ and $\tfrac{N_k}{2}.$
The initial state can be
given in the $J_{k,x}$ basis by \be \ket{\Psi_1}:= \sum_m f_1(m)
\ket{m}_{1,x}\ee and \be \ket{\Psi_2}:= \sum_m f_2(m) (-1)^m
\ket{m}_{2,x},\ee where for large particle numbers $f_m(x)\propto
\exp[-\tfrac{x^2}{N_m}].$ Similarly, we define for the
state of the light \be \ket{\Psi_l}:= \sum_m g(m) \ket{m}_{{\rm
light},z},\ee where $g(z)\propto \exp[-\tfrac{z^2}{2S_0}].$ Hence, we obtain the evolution
for the state of atoms and photons as
\begin{eqnarray}
\ket{\Phi(t)} &=&\exp(-iJ_xS_zt)
\ket{\Psi_1}\otimes\ket{\Psi_2}\otimes\ket{\Psi_l}\nonumber\\&=&
 \sum_{j_1,j_2}\sum_s e^{-i(j_1+j_2)s\Omega t} f_1(j_1)f_2(j_2)(-1)^{j_2}g(s) \ket{j_1}_x\ket{j_2}_x\ket{s}_{{\rm light},z}.
\label{integral2}
\end{eqnarray}
The projection to
the $S_y=0$ state can be incorporated into the model by introducing
$w_{s}=:\braket{S_z=s}{S_y=0}.$ Then, the final state of the atoms is
\begin{eqnarray}
\ket{\Psi(t)} &\propto&
 \sum_{j_1,j_2}G(j_1+j_2)f_1(j_1)f_2(j_2)(-1)^{j_2} \ket{j_1}_x\ket{j_2}_x,
 \label{integral}
\end{eqnarray}
where $G(j):=\sum_s e^{-ijs\Omega t} g(s) w(s).$ Here $G(j)$, for
large systems, is essentially the Fourier transform of $g(s)w(s).$
The value of $w(s)$ matters only for $\vert s \vert \lesssim
\sqrt{S_0}$ since for much larger $s$ we have $g(s)\approx 0.$
For this case, $w(s)$ is to a good approximation for successive
values of $s$ is alternating between $0$ and a constant. Hence,
$G(j)$ is very close to a Gaussian around $j=0$ with
a variance $\sim J(\tfrac{t}{\tau})^2.$ Thus,  $t\sim J^{0.25}\tau$
gives $\va{J_x}\sim \sqrt{J},$ while for $t\sim \tau_2:=\tau\sqrt{J}$ the width of
the Gaussian is $\sim1.$ That is, when computing $\ket{\Psi(t)}$, only states with $j_1+j_2=0$ are
selected
corresponding to projecting to the
$J_x=0$ subspace. Thus, for $t\sim \tau_2$ our setup realizes a von
Neumann measurement of $J_x.$ 
For large systems the summation in
\EQ{integral}
can be replaced by integration. Using these
ideas, we obtain \be\exs{J_x^2}\approx \int dj_1 dj_2 \bigg\vert
\braket{\Psi(t)}{j_1}_x\ket{j_2}_x\bigg\vert^2 (j_1+j_2)^2.\ee Due to the absolute
value sign, the $(-1)^{j_2}$ term in \EQ{integral} can be neglected for this
calculation, thus the integral of a smooth function must be computed
numerically for $\exs{J_x^2}.$
The dynamics of the other two variances can be computed similarly,
knowing the matrix elements of $M:=J_y^2+J_z^2.$
The change of $\exs{M}$ during this dynamics is negligible.
Note that $[J_x,M]=0,$ thus a von Neumann measurement of $J_x$
does not change $\exs{M}.$

\subsection{The $j>\tfrac{1}{2}$ case.}
Finally, the $j>\tfrac{1}{2}$ case is
analogous to the $j=\tfrac{1}{2}$ case, if we notice that starting
from a product state with $N_1$ particles in state $\ket{j}$ and
$N_2$ particles in state $\ket{-j}$ gives the same dynamics for
$\exs{J_{k,l}}$, as starting from a state with $2N_1j$ particles in
state $\ket{+\tfrac{1}{2}}$ and $2N_2j$ particles in state
$\ket{-\tfrac{1}{2}}.$

\end{document}